\documentclass[sigconf]{acmart}
\usepackage{pgfplots}
\usepackage{graphicx}
\usepackage{svg}

\newcommand{\x}{\mathbf{x}}
\newcommand{\z}{\mathbf{z}}

\newcommand{\n}{\mathbf{n}}

\newcommand{\xn}{\Tilde{\mathbf{x}}}

\newcommand{\encpar}{\phi_1}
\newcommand{\decpar}{\phi_2}
\newcommand{\clfpar}{\theta}
\newcommand{\encoder}[1]{e_{\encpar}\left(#1\right)}
\newcommand{\decoder}[1]{d_{\decpar}\left(#1\right)}
\newcommand{\advclf}[1]{a_{\clfpar}\left(#1\right)}
\newcommand{\preclf}[1]{c\left(#1\right)}

\DeclareMathOperator{\laplace}{Laplace}

\copyrightyear{2023}
\acmYear{2023}
\setcopyright{acmlicensed}
\acmConference[IH\&MMSec '23] {Proceedings of the 46th International ACM IH\&MMSec Conference on Research and Development in Information Retrieval}{June 28--30, 2023}{Chicago, IL, USA.}
\acmBooktitle{Proceedings of the 46th International ACM IH\&MMSec Conference on Research and Development in Information Retrieval (IH\&MMSec '23), June 28--30, 2023, Chicago, IL, USA}
\acmPrice{15.00}
\acmISBN{979-8-4007-0054-5/23/06}
\acmDOI{10.1145/3577163.3595102}

\begin{document}

\title{Differentially Private Adversarial Auto-Encoder\\to Protect Gender in Voice Biometrics}

\author{Oubaïda Chouchane}
\orcid{0000-0001-8208-9667}
\affiliation{%
  \institution{EURECOM}
  \city{Sophia Antipolis}
  \country{France}
}
\email{oubaida.chouchane@eurecom.fr}

\author{Michele Panariello }
\orcid{0009-0007-4154-5460}
\affiliation{%
  \institution{EURECOM}
  \city{Sophia Antipolis}
  \country{France}
}
\email{michele.panariello@eurecom.fr}

\author{Oualid Zari}
\email{oualid.zari@eurecom.fr}
\orcid{0009-0005-2093-4059}
\affiliation{%
  \institution{EURECOM}
  \city{Sophia Antipolis}
  \country{France}
}

\author{Ismet Kerenciler}
\orcid{0009-0009-6472-9208}
\affiliation{%
  \institution{EURECOM}
  \city{Sophia Antipolis}
  \country{France}
}
\email{Ismet.Kerenciler@eurecom.fr}

\author{Imen Chihaoui}
\orcid{0009-0008-7020-7062}
\affiliation{%
  \institution{EURECOM}
  \city{Sophia Antipolis}
  \country{France}
}
\email{Imen.Chihaoui@eurecom.fr}

\author{Massimiliano Todisco}
\orcid{0000-0003-2883-0324}
\affiliation{%
  \institution{EURECOM}
  \city{Sophia Antipolis}
  \country{France}
}
\email{massimiliano.todisco@eurecom.fr}

\author{Melek Önen }
\orcid{0000-0003-0269-9495}
\affiliation{%
  \institution{EURECOM}
  \city{Sophia Antipolis}
  \country{France}
}
\email{melek.onen@eurecom.fr}

\renewcommand{\shortauthors}{Oubaïda Chouchane et al.}

\begin{abstract}
Over the last decade, the use of Automatic Speaker Verification (ASV) systems has become increasingly widespread in response to the growing need for secure and efficient identity verification methods. The voice data encompasses a wealth of personal information, which includes but is not limited to gender, age, health condition, stress levels, and geographical and socio-cultural origins.
These attributes, known as soft biometrics, are private and the user may wish to keep them confidential. However, with the advancement of machine learning algorithms, soft biometrics can be inferred automatically, creating the potential for unauthorized use. As such, it is crucial to ensure the protection of these personal data that are inherent within the voice while retaining the utility of identity recognition. In this paper, we present an adversarial Auto-Encoder--based approach to hide gender-related information in speaker embeddings, while preserving their effectiveness for speaker verification. We use an adversarial procedure against a gender classifier and incorporate a layer based on the Laplace mechanism into the Auto-Encoder architecture. This layer adds Laplace noise for more robust gender concealment and ensures differential privacy guarantees during inference for the output speaker embeddings. Experiments conducted on the VoxCeleb dataset demonstrate that speaker verification tasks can be effectively carried out while concealing speaker gender and ensuring differential privacy guarantees; moreover, the intensity of the Laplace noise can be tuned to select the desired trade-off between privacy and utility.\\
\vspace{-.44cm}
\end{abstract}



\maketitle
\section{Introduction}
Voice is a unique biometric trait that is widely recognized for its capability to efficiently and securely identify individuals~\cite{sang2023improving}.
The use of voice as a biometric modality has been deployed in Automatic Speaker Verification (ASV) systems that have been incorporated into a range of applications like personal database access, credit card authorization, voice banking, and funds transfer. In speaker verification, also known as speaker authentication, a user claims their identity, and the system evaluates the truthfulness of that claim by comparing the speaker's biometric characteristics with the stored representation of the claimed identity. The system seeks to establish a match between the speaker's features and the claimed identity that surpasses a specified threshold. In instances where a match is not found, the speaker is rejected. 
Moreover, the voice does not only contain unique identity information but also physiological or psychological aspects like age, gender, emotions, accent, ethnicity, personality, and health condition, referred to as soft biometrics~\cite{soft1}, that can be detected automatically using machine learning systems~\cite{gender_age_emotion_detec}. 
The utilization of these soft biometric traits in conjunction with primary biometrics can provide additional information for the recognition process and improve recognition accuracy~\cite{rec_enhance_gender}. 
Studies in ~\cite{speech2face} also show that speakers' short recordings can be used to reconstruct their average-looking facial images that embody their physical characteristics such as age, gender, and ethnicity.
However, despite their potential use for legitimate processing purposes, soft biometrics are susceptible to malicious utilization. This can occur through unauthorized data processing that puts individuals at risk of privacy concerns such as discrimination, invasive advertising, extortion, and other forms of abuse. As a specific illustration, the finance sector has been shown to exhibit gender-based biases in loan provision~\cite{turkey_gender_disc}. 
This raises concerns regarding the potential existence of discriminatory lending practices that pose greater barriers to women than to men in the pursuit of starting a business enterprise~\cite{loan_gender_disc2}. Solutions based on cryptographic primitives~\cite{chouchane21,9794535}, while effective, produce completely garbled messages. Data obfuscation techniques, on the other hand, provide a more balanced approach to privacy preservation, protecting sensitive information without rendering the entire message content unrecognizable. Moreover, the voice is recognized as personal and sensitive and is therefore subject to protection under the General Data Protection Regulation (GDPR or Regulation 2016/679)\footnote{https://gdpr-info.eu/} together with numerous other data protection legislation, worldwide. The GDPR considers gender as well as a form of personal data and imposes an obligation to safeguard its protection.
In light of the increasing concerns surrounding privacy, there has been a growing effort to protect private information like soft biometrics. This effort has led to multiple research initiatives aimed at developing and implementing effective techniques for protecting the privacy of soft biometric attributes~\cite{terhorst2020pe,morales2020sensitivenets}.
Among these, techniques based on the differential privacy (DP) notion~\cite{DP} have received significant attention. Differentially private solutions (also referred to as global or centralized DP) were proposed for more than a decade and regarded as a privacy protection tool for different areas~\cite{US_Bureau, medical}. While global DP mechanisms consist of a trusted central party/data curator collecting the users' data, aggregating them, and further protecting the aggregated information by adding some calibrated noise before releasing it to the public, local DP (LDP) solutions ~\cite{LDP} protect the input data immediately to prevent the data curator from discovering the real, individual data. The noise is derived from a DP mechanism (e.g. Laplace mechanism).
In this paper, we aim to address the challenge of protecting gender information while preserving the efficiency of speaker verification. Our approach is based on adding a calibrated noise drawn from the Laplace distribution during the training of an Adversarial Auto-Encoder (AAE) architecture. The noise is injected into the latent space (i.e. the output of the encoder) in order to assure that the model is $\epsilon$-differentially private and to enhance the capability of the adversary in obscuring gender information. 
The speaker makes use of the private AAE locally to conceal their gender prior to the dissemination of their speaker features for the purpose of authentication.
Our experiments conducted on the VoxCeleb 1 and VoxCeleb 2 datasets demonstrate the feasibility of executing speaker verification tasks effectively while disrupting adversarial attempts of gender recognition.
To the best of our knowledge, this is the first work that uses differentially private solutions to protect gender information while preserving identity in biometrics.

\section{Related work}
In recent years, there has been a proliferation of academic literature pertaining to the topic of soft biometrics protection in biometric recognition systems. A significant number of researchers have centered their efforts on developing technical solutions that are capable of preventing the extraction of soft biometric attributes and are either directly applied to the collected biometric data like face images and voice signals (i.e. at sample level)~\cite{face2022gender, VC_gender_2021, aloufi2019emotionless, Semi-adversarial-Face2018} or to the extracted features (i.e. at feature level)~\cite{melzi2023multi,noe2020adversarial,bortolato2020learning, age_gender_suppressing_face2019}. 

Mirjalili et al.~\cite{Semi-adversarial-Face2018} proposed a Semi-Adversarial Network (SAN) based on an adversarial Convolutional Auto-Encoder (CAE) in order to hide the gender information from face images while retaining the biometric matching utility. In a follow-up work ~\cite{Semi-adversarial-Face2018gender}, the same authors introduced an ensemble of SANs that are constituted of multiple auxiliary gender classifiers and face matches that generates diverse perturbations for an input face image. The idea behind this approach is that at least one of the perturbed images succeeds in fooling an arbitrary gender classifier. In~\cite{mirjalili2019flowsan}, Mirjalili et al. also attempted to combine a variety of face perturbations in an effort to improve the generalization capability of SAN models. Despite the successful privacy preservation of gender attributes by the aforementioned techniques, their robustness to arbitrary classifiers is limited. In a more recent study, Tang et al.~\cite{face2022gender} presented an alternative gender adversarial network model that effectively masks gender attributes while preserving both image quality and matching performance. Besides, this model demonstrates the ability to generalize to previously unseen gender classifiers. Further work was proposed by Bortolato et al.~\cite{bortolato2020learning} to leverage the privacy-preservation of face images on the template level also using the AE technique. 
The authors suggested an AE-based solution that effectively separates gender attribute information from identity, resulting in good generalization performance across a variety of datasets. 
Additionally, Terhöst et al.~\cite{age_gender_suppressing_face2019} introduced an Incremental Variable Eliminations (IVE) algorithm that trains a set of decision trees to determine the importance of the variables that are crucial for predicting sensitive attributes. These variables were then incrementally removed from the facial templates to suppress gender and age features while maintaining high face-matching performance. In ~\cite{melzi2023multi} Melzi et. al. extended this approach to protect multiple soft biometrics (i.e. gender, age, and ethnicity) present in facial images. In speech-related literature, Aloufi et al.~\cite{aloufi2019emotionless} built a Voice Conversion (VC) system that can conceal the emotional state of the users while maintaining speech recognition utility for voice-controlled IoT. The model is based on a Cycle-Generative Adversarial Network (GAN) architecture.
Similarly, in~\cite{VC_gender_2021}, the authors introduced a neural VC architecture that can manipulate gender attributes present in the voice signal. This proposed VC architecture involves multiple Auto-Encoders that transform speech into independent linguistic and extra-linguistic representations. These representations are learned through an adversarial process and can be adjusted during VC.

On a template level, Noé et. al.~\cite{noe2020adversarial} proposed an adversarial Auto-Encoder architecture that disentangles gender attributes from x-vector speaker embeddings~\cite{x-vectors}. The AE is combined with an external gender classifier that attempts to predict the attribute class from the encoded representations. The proposed solution succeeds in concealing gender-related information in the embedding while maintaining good ASV performance. Nonetheless, our experimental findings indicate that using speaker embeddings other than x-vectors, such as those generated by the ECAPA-TDNN model~\cite{ecapa_is_cool_1}, yields inconsistent performance, implying potential challenges in achieving generalization.
We hypothesize that this may be attributed to the superior representational capabilities of ECAPA-TDNN embeddings, which have largely superseded x-vectors in recent speaker modeling.

\section{Gender concealment}
In this section, we present the building blocks of the proposed gender concealment technique.
First, we describe the architecture of the AAE and highlight its limitations in the concealment task.
Second, we briefly introduce local differential privacy, a concept that is instrumental in improving the gender concealment capabilities of the model.
Lastly, we illustrate how to combine the AAE and LDP to obtain a more effective technique for suppressing gender information in speaker embeddings, with a tunable privacy-utility trade-off and sound theoretical guarantees.

\subsection{Gender-Adversarial Auto-Encoder}
\label{sec:gaae}
Let $\x$ be an embedding representing a speaker identity. The goal of a Gender-Adversarial Auto-Encoder is to process $\x$ so as to produce a new embedding $\xn$ that still encodes the identity of that same speaker, but is devoid of any information about their gender. In this section, we describe our implementation of this system, which mostly follows the one proposed in~\cite{noe2020adversarial}.

Given an input embedding $\x \in \mathbb{R}^d$, we create a compressed representation of it by means of $\encoder{\x} = \z \in \mathbb{R}^l$, where $\encoder{\cdot}$ is a feed-forward neural network parameterized by $\encpar$ and $l < d$. The disentanglement of gender-related information from $\z$ depends on an adversarial ``discriminator'' module $\advclf{\cdot}$ (also a feed-forward neural network) that attempts to infer the gender of the speaker associated with $\z$. During training, we optimize $\clfpar$ to minimize the objective:
\begin{equation}
    \label{eq:discri_loss}
    \mathcal{L}_{disc}\left(\x, y, \clfpar \mid \encpar \right) = 
    - y \log\left(\advclf{\z}\right)
    - (1-y)\log\left(1-\advclf{\z}\right)
\end{equation}
where $y \in \left\{0,1\right\}$ is the ground-truth gender label ($0$ for male, $1$ for female) and $\advclf{\z} \in \left[0,1\right]$ represents the predicted probability of $\z$ having been produced by a female speaker.
The suppression of the gender-related information is performed by adversarially training the encoder to ``fool'' the discriminator, i.e. to make it so that it is not capable of accurately predicting the speaker's gender from $\z$. In practice, this is achieved by optimizing the same objective as \eqref{eq:discri_loss}, except that the probability predicted by the discriminator is inverted:
\begin{equation}
    \label{eq:adv_loss}
    \mathcal{L}_{adv}\left(\x, y, \encpar \mid \clfpar \right) = 
    - y \log\left(1 - \advclf{\z}\right)
    - (1-y)\log\left(\advclf{\z}\right)
\end{equation}
A decoder feed-forward module $\decoder{\cdot}$ attempts to reconstruct the original input embedding from $\z$.
The role of the decoder is to guarantee that the reconstructed embedding can still be used for other tasks, e.g. speaker verification, despite the suppression of gender-related attributes.
Thus, the Auto-Encoder is optimized end-to-end according to a further ``reconstruction'' objective: the cosine distance between the original input embedding and the reconstructed one.
\begin{equation}
    \label{eq:reco_loss}
    \mathcal{L}_{rec}\left(\x, \encpar, \decpar\right) =
    1 - \cos\left(\x, \decoder{\z}\right)
\end{equation}
Overall, we aim to strike a balance between privacy protection (optimizing $\mathcal{L}_{disc}$, $\mathcal{L}_{adv}$) and utility (optimizing $\mathcal{L}_{rec}$) of the processed embeddings. The overall system is trained by alternating gradient descent steps on the parameters of the Auto-Encoder $\phi = \{\encpar, \decpar\}$ and the parameters of the discriminator $\clfpar$:
\begin{equation}
    \begin{aligned}
    \phi & \leftarrow \nabla_{\phi} \left(\mathcal{L}_{adv} + \mathcal{L}_{rec}\right)\\
    \clfpar & \leftarrow \nabla_{\clfpar} \mathcal{L}_{disc}
    \end{aligned}
\end{equation}

At test time, we produce a protected embedding $\xn$ by passing $\x$ through the Auto-Encoder:
\begin{equation}
    \xn = \decoder{\encoder{\x}}
\end{equation}
The privacy preservation capability of the Auto-Encoder is evaluated upon the ability of an attacker to infer the gender of the original speaker from the protected utterance $\xn$.
To measure it, we train an external gender classifier $\preclf{\cdot}$ on a separate set of clean embeddings, then report the gender classification performance of $\preclf{\cdot}$ on the original test embeddings and their privacy-protected version:
the difference between the two represents the effectiveness of gender concealment.
The utility preservation is evaluated by comparing the performance of the same ASV system on the original and protected speaker embeddings.

We perform a preliminary evaluation of the reconstructed speaker embeddings of the Gender-AAE and obtain Area Under the ROC Curve (AUC) for gender recognition  = 98.45 ($10^{-2}$) and Equal Error Rate (EER) = 1.86\% for ASV performance. In order to ensure that the predictions of the gender classifier are truly random, the AUC must be close to 50\%. Therefore, it is necessary to strengthen the adversarial performance to conceal gender information. 

In this work, we investigate the impact of adding noise derived from a Laplace mechanism which is well-studied for noise addition and calibration and also provides DP guarantees. The latent vectors $\z$ are locally differentially private thanks to the Laplace mechanism and subsequently, the reconstructed vectors are differentially private by the post-processing property of DP~\cite{post-proc}. 

\begin{figure}[t]
     \centering
     \includegraphics[width=\columnwidth]{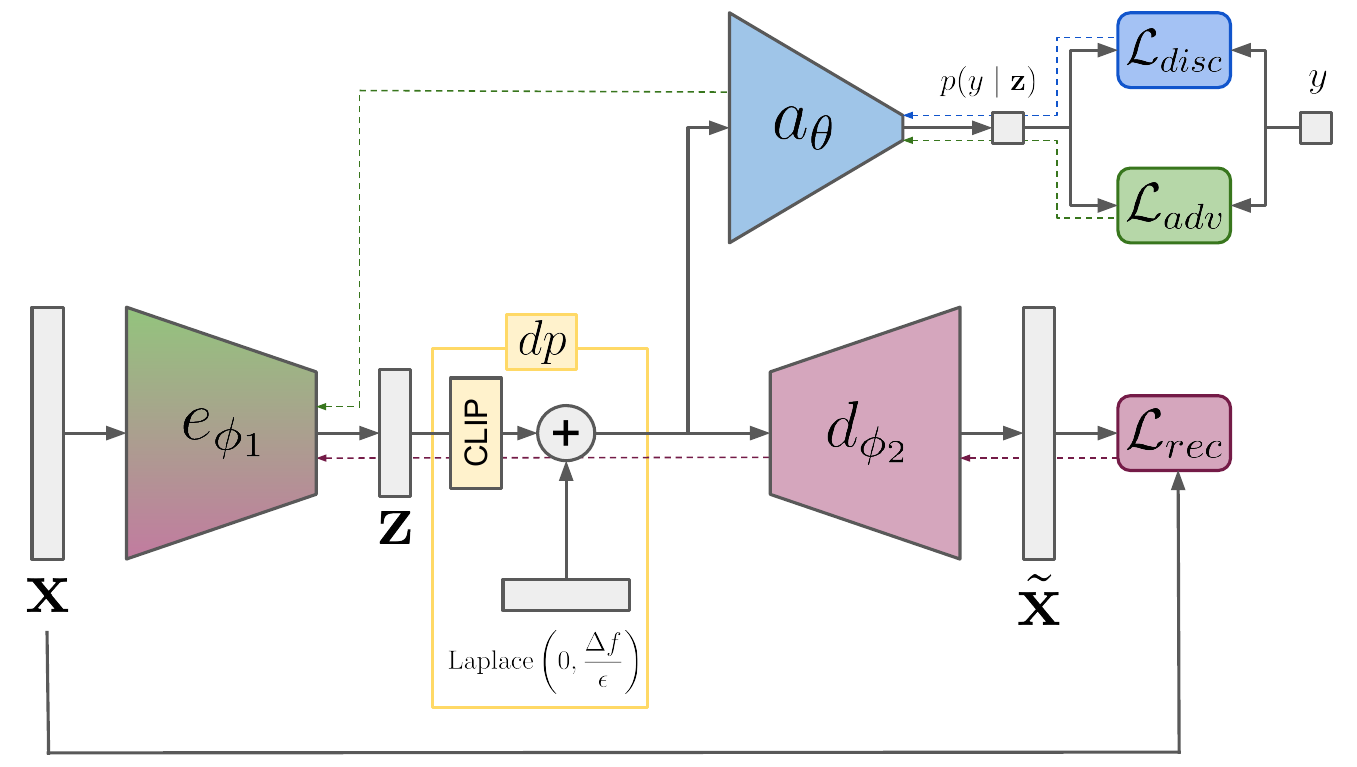}
     \caption{Illustration of the proposed system at training time. Solid and dashed arrows represent forward and backward propagation respectively. Modules are colored based on which gradient signal they are optimized by. }
     \vspace{-0.5cm}
     \label{fig:aae}
\end{figure}

\subsection{Local Differential Privacy}
Local differential privacy plays a crucial role in protecting personal data like soft biometrics and assessing the privacy risks. In this section, we provide a brief description of the underlying concepts of local differential privacy and the Laplace mechanism.

\textbf{\textit{Definition.}}
Local differential privacy is a state-of-the-art privacy model and consists in protecting individual input data before its collection.
LDP ensures privacy for each user locally (i.e. each individual record is protected rather than the entire dataset as a whole) by adding noise without the necessity of trusting a central authority. 
Formally, ($\epsilon$)-local differential privacy is defined as follows. 
\begin{definition}[(Local Differential Privacy~\cite{LDP}]
A randomized algorithm $\mathcal{M}$ satisfies ($\epsilon$)-LDP if and only if for any pairs of input values $x, x' \in \mathcal{X}$ in the domain of $\mathcal{M}$, and for all possible outputs $S\subseteq Range(\mathcal{M})$, we have:
\begin{equation}
Pr[\mathcal{M}(x) \in S] \leq e^\epsilon \cdot Pr[\mathcal{M}(x') \in S]
\end{equation}
\end{definition}
where $Pr$ denotes the probability and $\epsilon$ ($\epsilon >0$) is known as the privacy budget that provides a measure of the privacy loss incurred by the DP algorithm. The smaller the value of $\epsilon$, the smaller the privacy loss (i.e. the stronger the privacy protection) and vice versa.

\textbf{\textit{Sensitivity.}}
The sensitivity~\cite{sensitivity}, denoted as $\Delta f$, is a measure of the maximum influence that a single data point can have on the result of a numeric query $f$. In an LDP mechanism, the sensitivity can be defined as shown in \eqref{eq_sensitivity}, where $x$ and $x'$ represent two adjacent records in a dataset $\mathcal{X}$ and $\lVert . \rVert$ denotes the $\ell_1$ norm of a vector.
\begin{equation} 
\label{eq_sensitivity}
\Delta f = \max_{x,x' \in\mathcal{X}} \left \| f(x) - f(x')\right \|_1
\end{equation}

The sensitivity is the maximum difference between two adjacent records in a dataset and it provides an upper bound on the potential impact of an individual record. It defines the magnitude of the noise needed in order to meet the ($\epsilon$)-LDP requirements. 

\textbf{\textit{Laplace Mechanism.}}
The Laplace mechanism~\cite{laplace} is a widely adopted technique for achieving ($\epsilon$)-LDP. The mechanism works by adding random noise, sampled from the Laplace distribution, to the output of a function in order to obscure any sensitive information about individual records in the database. The amount of noise added is determined by the sensitivity $\Delta f$ of the function and the privacy budget $\epsilon$.
Formally, given a database $\mathcal{X}$ and a function $f:\mathcal{X} \rightarrow \mathbb{R}^{d}$ that maps the database to $d$ real numbers, the Laplace mechanism is defined as:
\begin{equation}
    \mathcal{M}(f(x),\epsilon) = f(x) + (n_1,n_2, ...,n_d).
\end{equation}
where each $n_i \sim \laplace\left(\Delta f / \epsilon\right)$ is drawn from the zero centered Laplace distribution with scale $\Delta f / \epsilon$.

The Laplace mechanism has been demonstrated to be particularly effective in the context of numerical queries (e.g. counting queries, histogram queries, and classification queries) with low sensitivity~\cite{DP}.
In our work, we use the Laplace mechanism to perturb each component of the latent speaker embedding $\z$ with noise drawn from the Laplace distribution. This approach successfully conceals the speaker's gender while retaining the usefulness of the feature vectors for ASV tasks.
\vspace{-.2cm}
\subsection{Gender-Adversarial Auto-Encoder  with Laplace noise}
We improve the gender concealment capability of the AAE by applying the Laplace mechanism to the latent space learned by the encoder.
More specifically, during training, we pass the latent embedding $\z$ through a \emph{Laplace} layer $dp\left(\cdot\right)$ that adds to its input a noisy vector $\n \sim \laplace\left(0, \Delta f / \epsilon\right)$. Figure~\ref{fig:aae} graphically depicts the system.

As there is no prior bound on the $\ell_1$-norm of the vector $\z$, we use the same clipping procedure described in \cite{dpsgd}: it consists in scaling $\z$ by a coefficient $1 / \max\left(1, \|\z\|_1 / C\right)$, where $C$ is the clipping threshold. This method ensures that if $\|z\|_1 \leq C$, $\z$ remains unchanged, while if $\|z\|_1 > C$, it is scaled down to have a norm of $C$. The purpose of the clipping is to ensure that the sensitivity between any pair of vectors $\z$ and $\z'$ is $\Delta f \leq 2C$.
In practice, one pragmatic approach to determine an appropriate value for $C$ is to compute the median of the norm of unclipped $\z$ vectors throughout the training phase.
Thus, the Laplace layer is defined as
\begin{equation}
    dp\left(\z\right) = \frac{\z}{\max\left(1, \frac{\|\z\|_1}{C}\right)} + \n
\end{equation}
and has no learnable parameters. It is applied before $\z$ is passed to the decoder $\decoder{\cdot}$ and to the discriminator $\advclf{\cdot}$.
The rest of the forward pass, the loss computation, and the overall training method then proceed as reported in Section~\ref{sec:gaae}.
Once the model has been trained, the adversarial module $\advclf{\cdot}$ is removed.
The value of $\epsilon$ can be chosen according to the desired balance between privacy protection and the utility of the produced embeddings.

The goal of adding Laplace noise into the system is twofold. At test time, its purpose is to provide privacy protection theoretical guarantees as previously described; however, at training time, it also serves as a regularizer for the adversarial module and the decoder. Indeed, our experiments show that applying the Laplace mechanism at training time only (i.e. removing the Laplace layer at test time) is sufficient to greatly enhance the gender concealment capabilities of the system.

To better explore the functional difference between the Laplace noise at test and at training time, we perform experiments by independently varying the value of $\epsilon$ during the training phase ($\epsilon_{tr}$) and during the testing phase ($\epsilon_{ts}$).
Our results show that changing $\epsilon_{tr}$ is the most convenient way to roughly set the balance between the empirical capabilities of gender concealment and ASV performance; however, by definition, $\epsilon_{tr}$ does not provide full control over the DP budget of the embeddings at test time.
Changing $\epsilon_{ts}$ then offers a flexible means of fine-tuning the privacy budget of the embeddings even once the model has been trained and deployed.

In Section~\ref{sec:experiments}, we show that both $\epsilon_{tr}$ and $\epsilon_{ts}$ are equally relevant in determining the behavior of the system.

\textbf{\textit{Privacy Guarantees for the Gender-AAE.}}
One of the main strengths of differential privacy lies in its property of post-processing, which ensures that the privacy guarantee offered by a DP mechanism remains unaltered regardless of the arbitrary computations performed on its output.
More formally, 
\vspace{-.2cm}
 \begin{definition}[Post-processing~\cite{laplace,post-proc}]
   Let $\mathcal{M}$ be an $\epsilon$-differentially private mechanism and $g$ be an arbitrary mapping from the set of possible outputs to an arbitrary set. Then, $g \circ \mathcal{M}$ is $\epsilon$-differentially private.  
 \end{definition}
 \vspace{-.2cm}
Similarly to the work in ~\cite{lecuyer2019certified}, we add noise to the latent space of the Auto-Encoder during the training, and use the same privacy proof, thanks to the post-processing property: 
$d_{\decpar}\circ dp$ satisfies $\epsilon$-DP, and so does the Auto-Encoder $d_{\decpar} \circ dp \circ e_{\encpar}$.

\begin{figure}[t]
    \centering
    \includegraphics[width=\columnwidth]{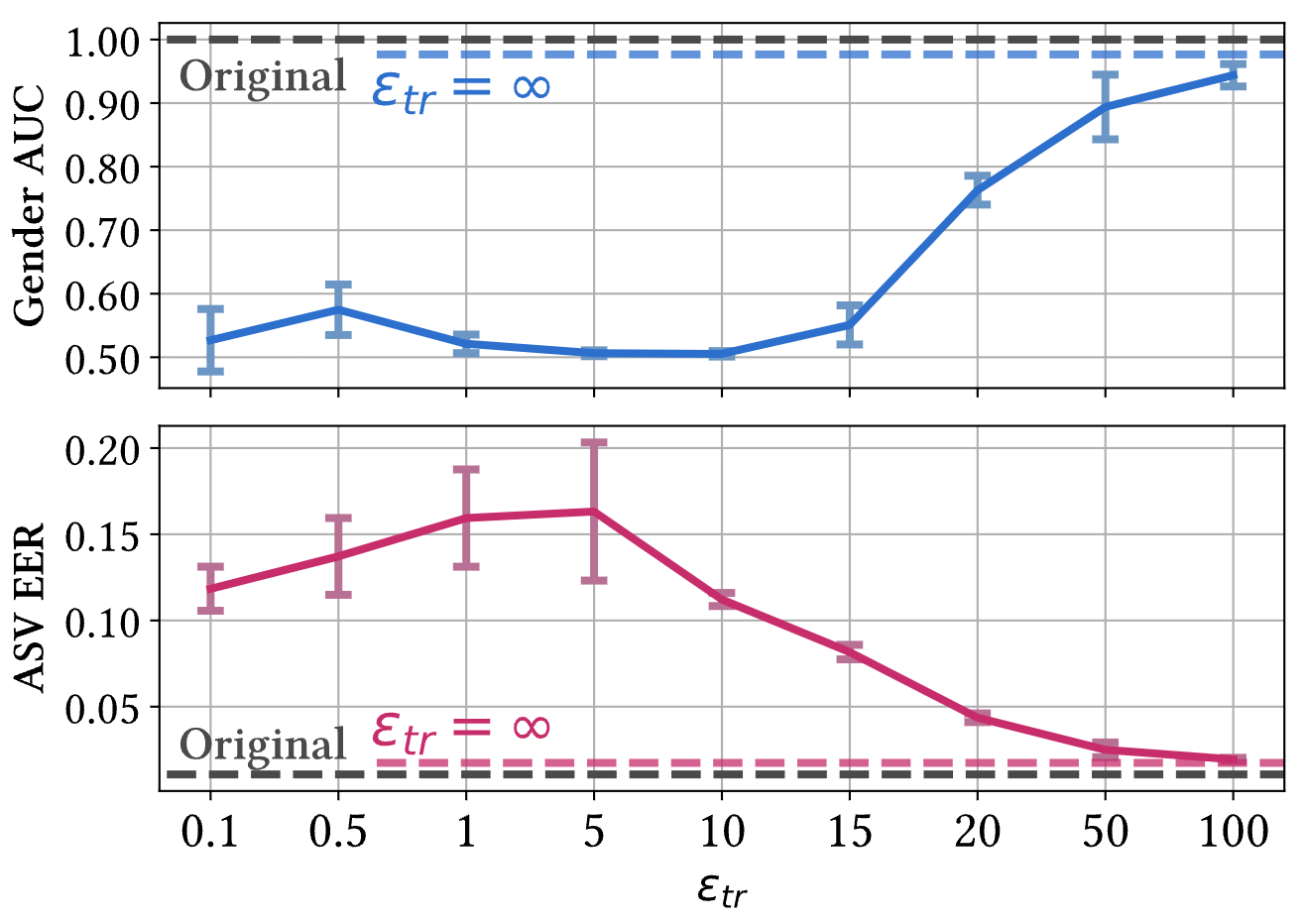}
    \caption{ASV EER and gender classification AUC achieved by the system for increasing values of $\epsilon_{tr}$.}
    \label{fig:err_auc}
    \vspace{-.4cm}
\end{figure}

\section{Experimental setup and results}
\label{sec:experiments}
In this section, we discuss the experimental configurations and results.
The feature extractor used to produce the speaker embeddings is the ECAPA-TDNN, whose output feature size is $d=192$.
The modules of the proposed encoder and decoder models are single-layer fully-connected neural networks and the gender classifiers (i.e. discriminator and external) are two-layer fully-connected neural networks.
The encoder is followed by a ReLU activation and batch normalization, and the decoder is followed by a tanh activation function. We set the latent space to be of size $l=64$.
The adversarial classifier is composed of two fully-connected layers: the first one has 64 input units with a ReLU activation function, and the second one has 32 input units with a sigmoid activation function.
An external gender classifier, used by an attacker to infer gender, is used to assess privacy protection and has the same architecture as the discriminator with 192 input units in the first layer and 100 input units in the second layer. The ASV assessment is done by first creating a model for each speaker; trial scores are then obtained by comparing trial embeddings with the respective speaker models by means of cosine similarity. The training process is carried out with Adam optimizer using a learning rate of $1 \cdot 10^{-3}$ and a minibatch size of 128.
The training dataset of the AAE is a subset of VoxCeleb2~\cite{voxceleb2} development partition (397032 segments per class). The testing is conducted using a subset of the VoxCeleb1~\cite{voxceleb1} test partition (2900 segments per class). The external sex classifier is trained using a subset of the VoxCeleb1 development partition (61616 segments per class).
To select the clipping threshold $C$, we compute the median of the norm of all unclipped $z$ vectors during the training, which is $C$=18.35.

We initially explore the behavior of the system by setting $\epsilon_{ts}=\infty$ (i.e. no DP protection) and for increasing values of $\epsilon_{tr}$: Figure~\ref{fig:err_auc} shows the achieved ASV EER and gender classification AUC.
We experimentally determine the noise scale and prioritize higher $\epsilon_{tr}$ resolution for the region with significant privacy/utility changes, while lower resolution suffices for regions with minor variations.
As expected, privacy and utility scores inversely mirror one another.
Specifically, $\epsilon_{tr} = 15$ seems to strike a satisfactory balance between the two, resulting in a 0.55 gender classification AUC while achieving an ASV EER of 8.1\%. For comparison, the same gender classifier and ASV system obtain an AUC of nearly $1$ and an EER of $1.1\%$ on the original ECAPA embeddings, respectively.

We pick the model weights trained with $\epsilon_{tr} = 15$ and $\epsilon_{tr} = 20$ and experiment with values of $\epsilon_{ts} < \infty$ to add DP protection to the speaker embeddings.
Setting $\epsilon_{ts} = \epsilon_{tr}$ further enhances the level of gender concealment: AUC scores drop from 0.55 to 0.50 (from 0.76 to 0.55 respectively) for $\epsilon_{tr} = \epsilon_{ts} = 15$ ($\epsilon_{tr} = \epsilon_{ts} = 20$ respectively). However, ASV EER degrades by around 20 percentage points in both scenarios.
By increasing $\epsilon_{ts}$ by 20 units, it is possible to restore the ASV EER to around 10\% (for both model versions) while achieving satisfactory AUC values of 0.55 and 0.68 for $\epsilon_{tr}=15$ and $\epsilon_{tr}=20$, respectively.
In general, these results show the level of flexibility that the system can achieve even after training, all while providing DP guarantees over the produced embeddings.

Informal experiments run with $\epsilon_{tr}=\infty$ have resulted in rapid erasure of all meaningful information from the speaker embeddings even for high values of $\epsilon_{ts}$: this is indicative of the relevance of including the Laplace noise during training for the DP protection to be applicable at test time.

\begin{figure}[t]
    \centering
    \includegraphics[width=\columnwidth]{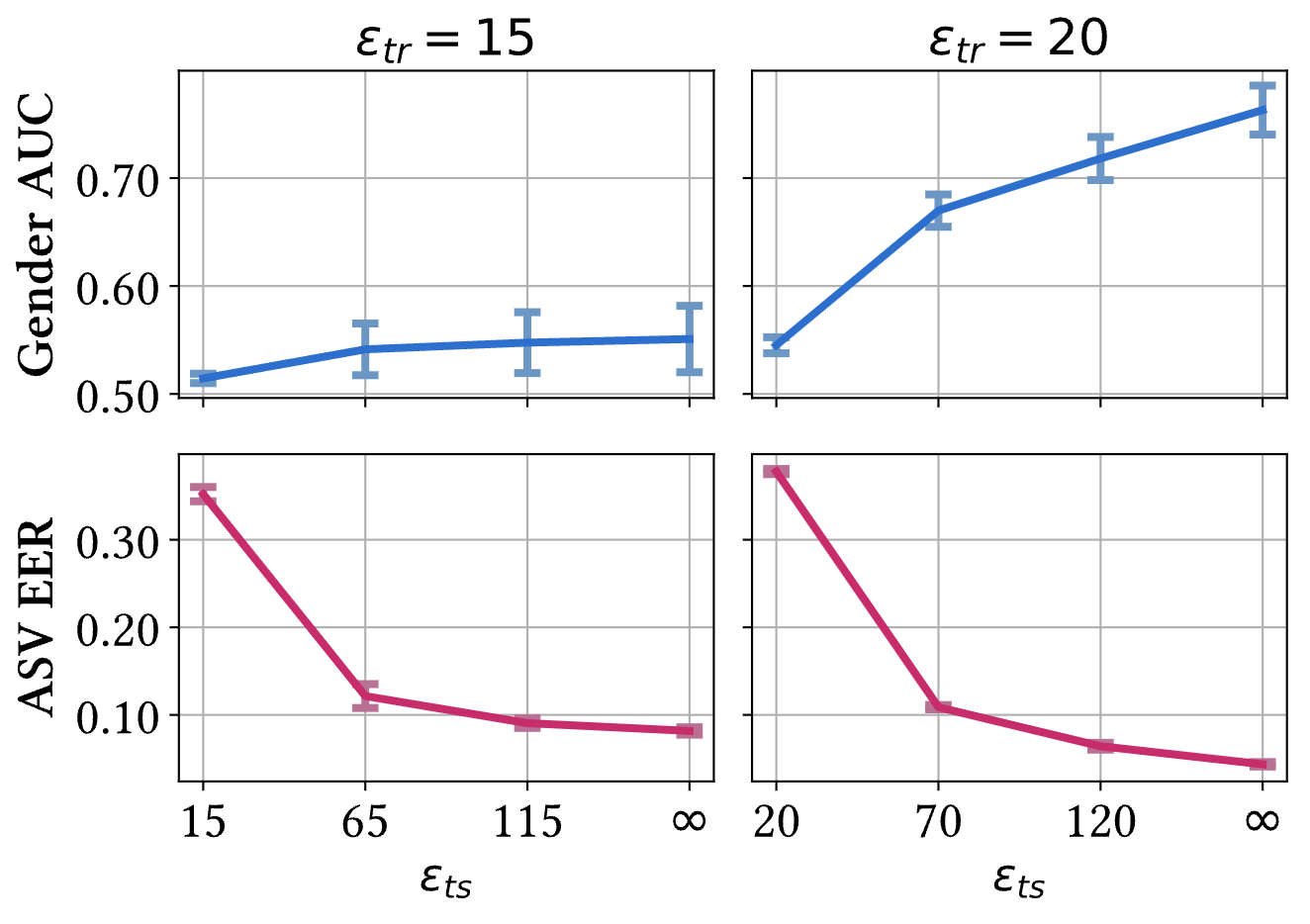}
    \caption{ASV EER and gender classification AUC achieved by the system for increasing values of $\epsilon_{ts}$, for the cases of $\epsilon_{tr}=15$ and $\epsilon_{tr}=20$.}
    \label{fig:err_auc_test}
    \vspace{-.4cm}
\end{figure}

\section{Conclusions}
We have presented an AE-based system to conceal gender-related information in speaker embeddings while retaining their utility for a speaker verification task.
We perform the concealment by means of an adversarial game between an Auto-Encoder and an external gender classifier, and we improve upon previous work by introducing a Laplace-noise--addition layer within the architecture. The Laplace noise regularizes the training and allows for more robust gender concealment, while also endowing the output speaker embedding with DP guarantees at inference time.
The tuning of the $\epsilon$ parameter of the Laplace layer allows selecting the desired balance of privacy protection and utility, even after the training process has finished.
Experimental results show that the proposed solution is effective in preserving gender privacy while maintaining utility for speaker verification tasks. Furthermore, the flexible trade-off between privacy and utility provided by our approach can be adapted to individual needs, making it a promising solution for privacy-preserving applications.

\begin{acks}
This work is supported by the TReSPAsS-ETN project funded by the European Union’s Horizon 2020 research and innovation programme under the Marie Skłodowska-Curie grant agreement No. 860813. It is also supported by the ANR-DFG RESPECT project.
\end{acks}

\bibliographystyle{ACM-Reference-Format}
\balance
\bibliography{biblio}

\end{document}